\documentclass{PoS}

\title{Constrained Supersymmetry after two years of LHC data: a global view with Fittino}

\ShortTitle{Constrained Supersymmetry after two years of LHC data: a global view with Fittino}

\author{Philip Bechtle,$^a$ Torsten Bringmann,$^b$ Klaus Desch,$^a$ Herbi Dreiner,$^{ac}$
Matthias Hamer,$^d$ Carsten Hensel,$^d$ Michael Kr\"amer,$^e$ Nelly Nguyen,$^f$ Werner Porod,$^g$
\speaker{Xavier Prudent},$^h$ Bj\"orn Sarrazin,$^i$ Mathias Uhlenbrock $^a$ and Peter Wienemann $^a$.
\llap{$^a$}Physikalisches Institut, University of Bonn,\\
 Bonn, Germany\\
\llap{$^b$}II. Institute for Theoretical Physics, University of Hamburg,\\
Hamburg, Germany\\
\llap{$^c$}Bethe Center for Theoretical Physics, University of Bonn,\\
Bonn, Germany\\
\llap{$^d$}II. Physikalisches Institut, University of G\"ottingen,\\
G\"ottingen, Germany\\
\llap{$^e$}Institute for Theoretical Particle Physics and Cosmology, RWTH Aachen University,\\
Aachen, Germany\\
\llap{$^f$}Institute for Experimental Physics, University of Hamburg,\\
Hamburg, Germany\\
\llap{$^g$}Institut f\"ur Theoretische Physik und Astrophysik, University of W\"urzburg,\\
W\"urzburg, Germany\\
\llap{$^h$}Institut f\"ur Kern- und Teilchenphysik, TU Dresden,\\
Dresden, Germany\\
\llap{$^i$}Deutsches Elektronen-Synchrotron DESY,\\
Hamburg, Germany\\
E-mail: \email{bechtle@physik.uni-bonn.de}, \email{torsten.bringmann@desy.de},
\email{desch@physik.uni-bonn.de}, \email{dreiner@th.physik.uni-bonn.de},
\email{mhamer@uni-goettingen.de}, \email{carsten.hensel@cern.ch},
\email{mkraemer@physik.rwth-aachen.de}, \email{nelly.nguyen@desy.de},
\email{porod@physik.uni-wuerzburg.de}, \email{prudent@physik.tu-dresden.de},
\email{bjoern.sarrazin@desy.de}, \email{Mathias.Uhlenbrock@cern.ch},
\email{wienemann@physik.uni-bonn.de}}

\abstract{We perform global fits to the parameters of the Constrained Minimal Supersymmetric Standard Model (CMSSM) and to a variant with non-universal Higgs masses (NUHM1). In addition to constraints from low-energy precision observables and the cosmological dark matter density, we take into account the LHC exclusions from searches in jets plus missing transverse energy signatures with about 5\,fb$^{-1}$ of integrated luminosity at $\sqrt s=7~$TeV. We also include the most recent upper bound on the branching ratio $B_s\to\mu\mu$ from LHCb. The best fit of the CMSSM prefers a light Higgs boson just above the experimentally excluded mass. We find that the description of the low-energy observables, $(g-2)_{\mu}$ in particular, and the non-observation of SUSY at the LHC become more and more incompatible within the CMSSM. A potential SM-like Higgs boson with mass around 126 GeV can barely be accommodated. Values for $BF(B_s\to\mu\mu)$ just around the Standard Model prediction are naturally expected in the best fit region.}

\FullConference{36th International Conference on High Energy Physics\\
		 4-11 July 2012\\
		 Melbourne, Australia}

\begin{document}

\section{Introduction}

Current data being insufficient to constraint a general supersymmetric model, the most widely considered constrained model is the constrained MSSM (CMSSM) with only 5 new free
parameters beyond the SM: $M_0, M_{1/2}, A_0, \tan\beta, sgn(\mu)$, denoting respectively the universal soft supersymmetry breaking scalar and gaugino
masses at the unification scale, the universal soft supersymmetry breaking
trilinear scalar coupling, the ratio of the vacuum expectation values of the two
CP-even neutral Higgs fields and the Higgs mixing parameter in the superpotential.
In a minimal non-universal Higgs mass model (NUHM1)~\cite{NUHM} a universal scalar Higgs mass parameter $M_H$
at the unification scale, $M_{Hu} = M_{Hd} = M_H$, is added. In this paper, we investigate the global interpretation of all existing data using our framework Fittino~\cite{Fittino}.

\section{The Fittino framework}

The Fittino~\cite{Fittino,Fittino2} framework is used to perform a global Markov Chain Monte Carlo (MCMC) scan of the supersymmetric parameter space in all parameter dimensions. In the frequentist interpretation of the MCMC fit, assuming a Gaussian likelihood the two-dimensional $1\sigma$ ($2\sigma$) boundaries are defined by $\Delta\chi^2 < 2.33 (5.99)$, where $\Delta\chi^2$ is calculated for each point by regard to the best fit point with the smallest $\chi^2$. Using a self-optimizing chain, at least 3 million points are obtained within $\Delta\chi^2 < 5.99$ from the minimum for each individual fit.
The SUSY particle spectrum is calculated with SPheno 3.1.4~\cite{SPheno}, and then used in micrOMEGAs 2.2~\cite{Micromegas} for the prediction of the dark matter relic density, in FeynHiggs 2.8.2~\cite{FeynHiggs} for the prediction of the electroweak observables, for the Higgs masses and for the anomalous magnetic moment of the muon $a_{\mu}$, in SuperISO 3.1~\cite{SuperIso} for the flavor physics observables, and in AstroFit~\cite{AF} for the evaluation of the direct and indirect detection of dark matter observables. 

\section{Experimental constraints}

\begin{figure}
\includegraphics[width=.45\textwidth]{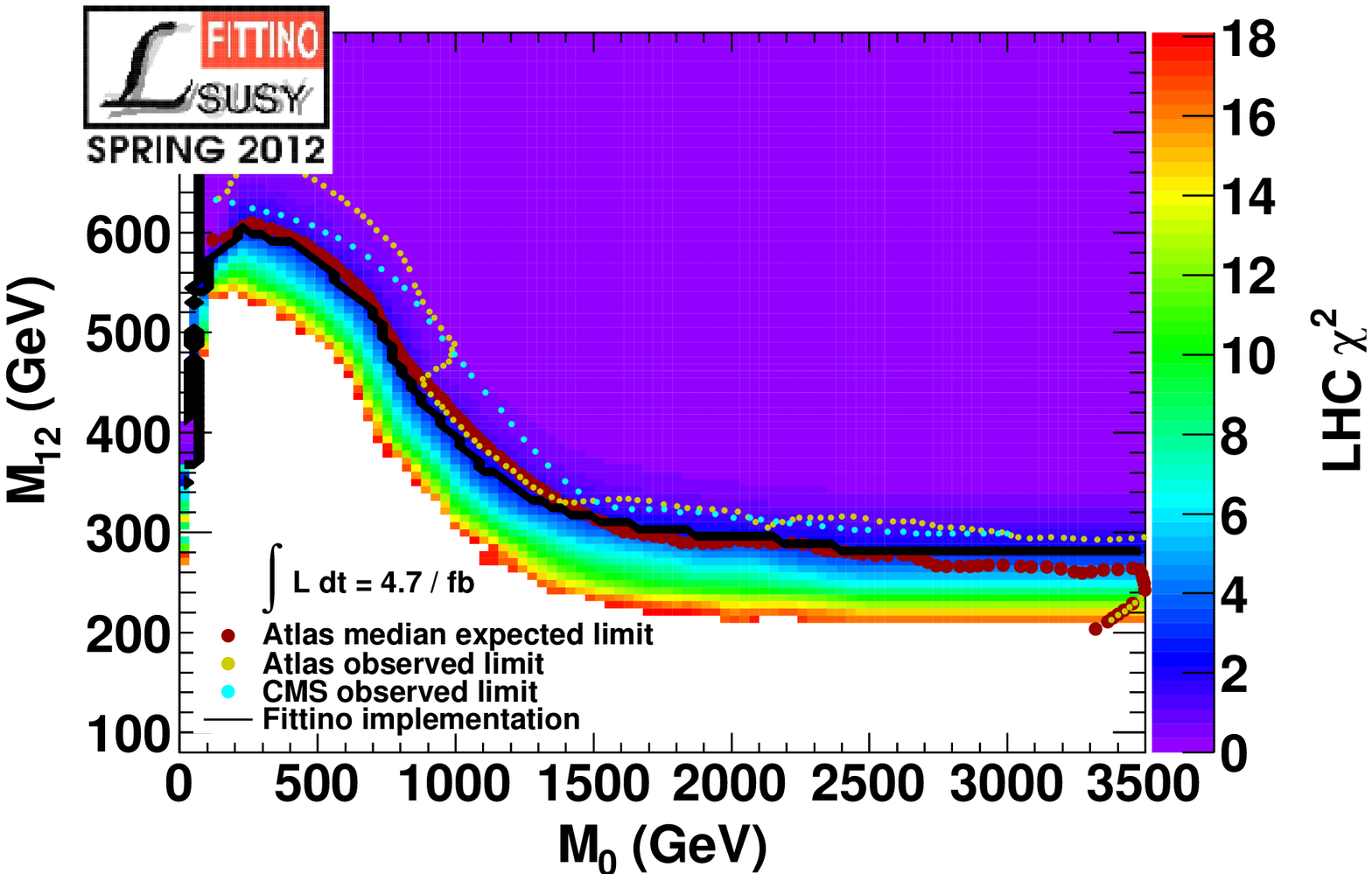}
\includegraphics[width=.45\textwidth]{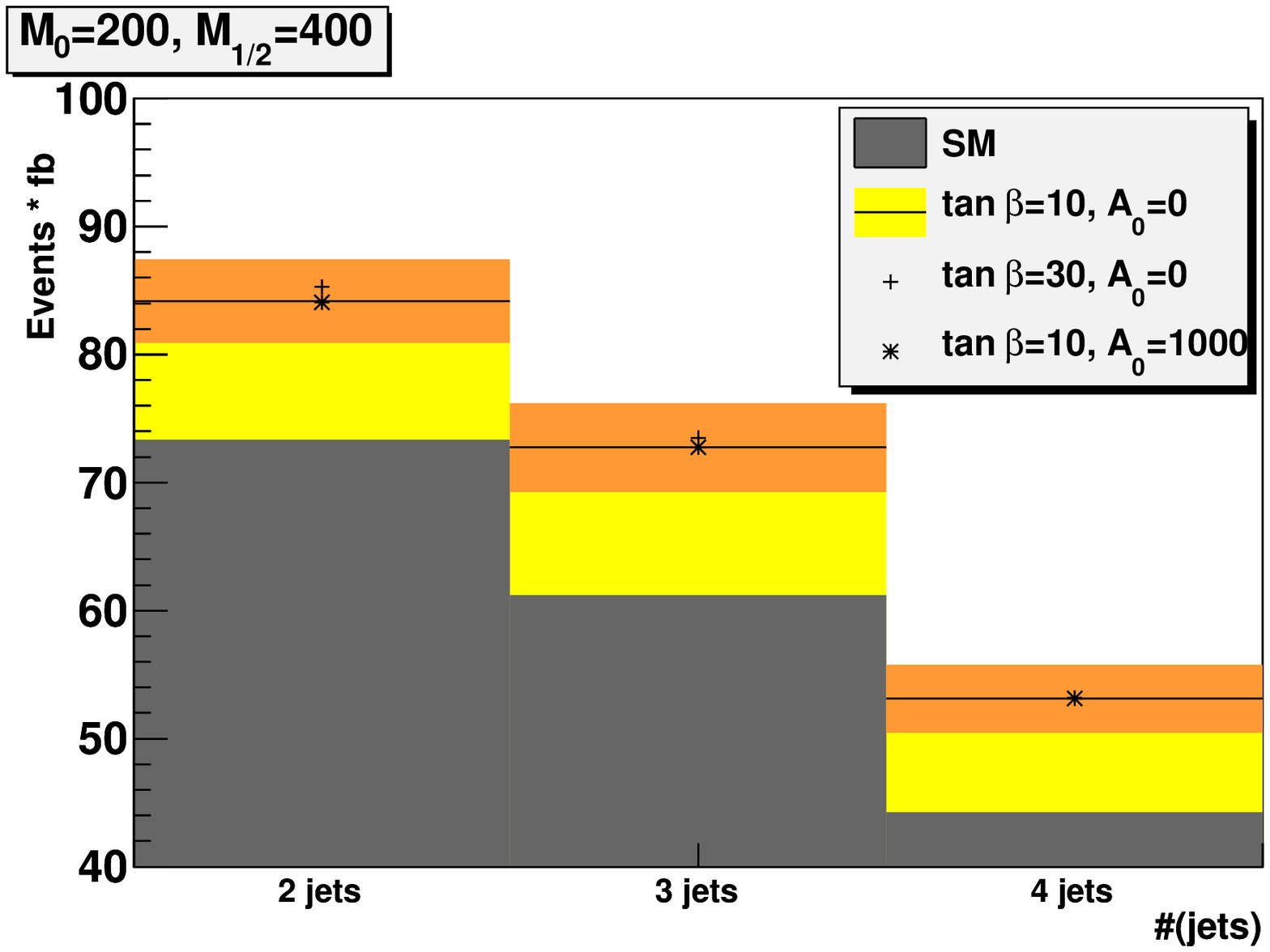}
\caption{Left: $\chi^2$ contribution from the LHC SUSY search implementation compared to the published ATLAS and CMS limits~\cite{ATLASAn}~\cite{CMSAn}. Right: Simulated signal yields for a different point in the $(M_0,M_{1/2})$ parameter space, with the SM background (gray), the CMSSM signal expectation for our grid (yellow) and the systematic uncertainty of 30$\%$ (orange).}
\label{fig:LHCgrid}
\end{figure}

A detailled description of the experimental constraints can be found in~\cite{mainPaper}.
The present and potential experimental measurements used in this study are indirect constraints through supersymmetric loop corrections and constraints from astrophysical observations.
The available limits on SM and non-SM Higgs bosons, including the ones presented by the LHC and Tevatron collaborations at the Spring Conferences 2011, are evaluated using HiggsBounds 3.2~\cite{HB}. An extensive database of relevant astrophysical data are added by Astrofit to the Fittino fit process. The most conservative chargino mass limit of $m_{\chi^{\pm}_1}>102.5$~GeV, from the direct search at LEP was included, leading indirectly to the exclusion of light neutralinos $m_{\chi_1^0}<50$~GeV in the constrained models considered here~\cite{ChargNeut}.
The most stringent LHC limits from the direct searches in channels with jets and missing transverse energy are included by emulating the search analysis. The ATLAS analysis~\cite{ATLASAn} is reproduced by simulating the production of gluino and squarks with SPheno 3.1.0 and Herwig 2.4.2~\cite{HERWIG}, together with the fast detector simulation DELPHES 1.9~\cite{DELPHES}. It is indeed not sufficient to only consider the $95\%$CL bounds published by the experimental collaborations for specific models and particular choices of parameters. The published ATLAS limits at $L_{int} = 165 /pb$ and $L_{int} = 1 /fb$ could be precisely reproduced (see Fig.~\ref{fig:LHCgrid}), the signal grid was adapted to the ATLAS analysis for $L_{int} = 4.7 /fb$ by reducing the systematic uncertainties. The impact of fixing $A_0$ and $\tan\beta$ on the grid was checked by simulating the signal for various values of $A_0$ and $\tan\beta$ in different regions of the $(M_0,M_{1/2})$ parameter space, and was found to be negligible (see Fig.~\ref{fig:LHCgrid}).

\section{Results}

\begin{figure}
\includegraphics[width=.45\textwidth]{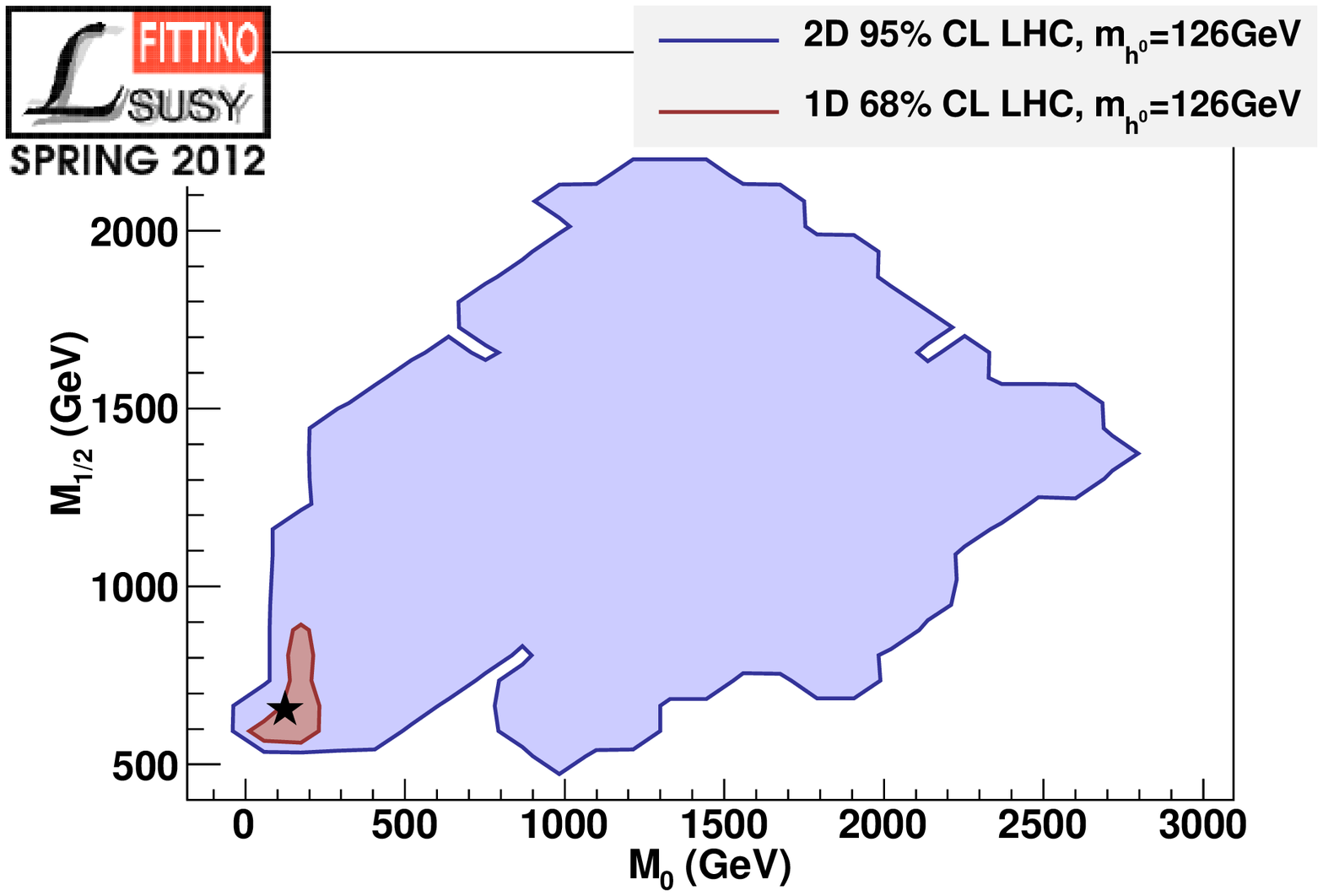}
\includegraphics[width=.45\textwidth]{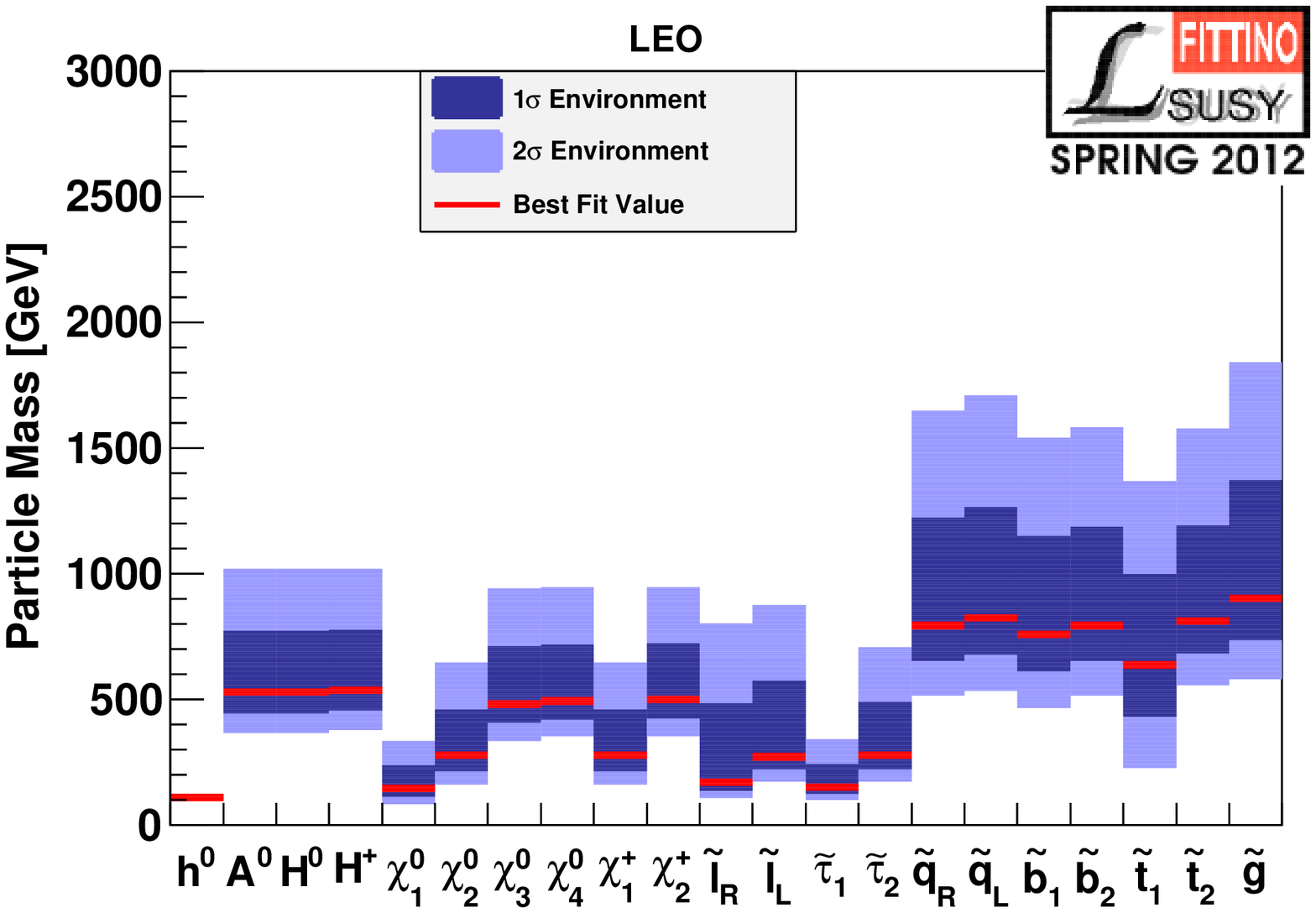}
\caption{Left: Parameter distributions for the LEO fit of the CMSSM with the 1-dimensional $1\sigma$ in red and the 2-dimensional $2\sigma$ in blue, and the best fit point marked by a star.
Right: predicted distribution of sparticle and Higgs boson masses from the LEO fit of the CMSSM. The full uncertainty band gives the 1-dimensional $2\sigma$ uncertainty of each mass $\Delta\chi^2 < 4$.}
\label{fig:LEO}
\end{figure}

\begin{figure}
\includegraphics[width=.45\textwidth]{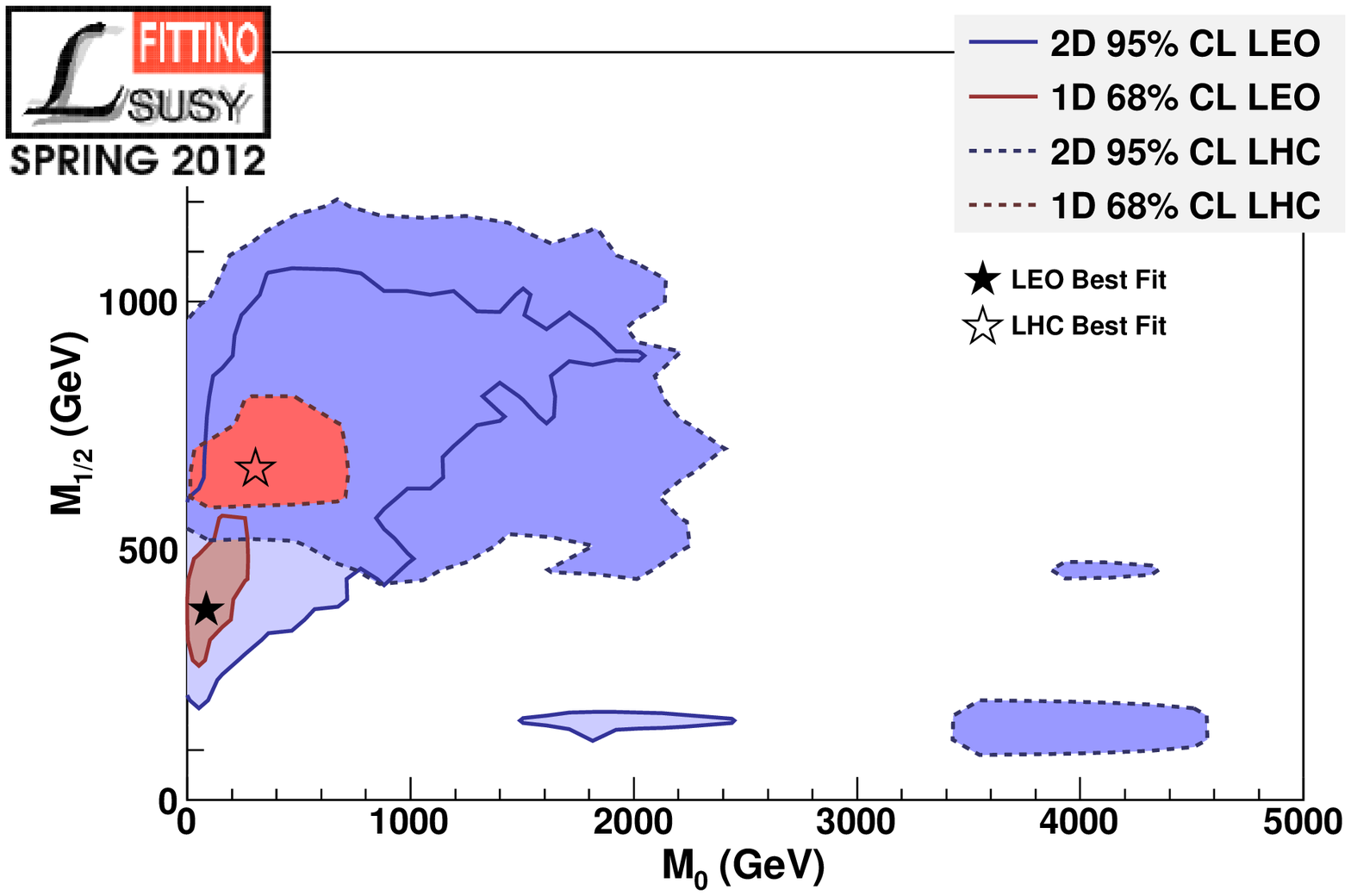}
\includegraphics[width=.45\textwidth]{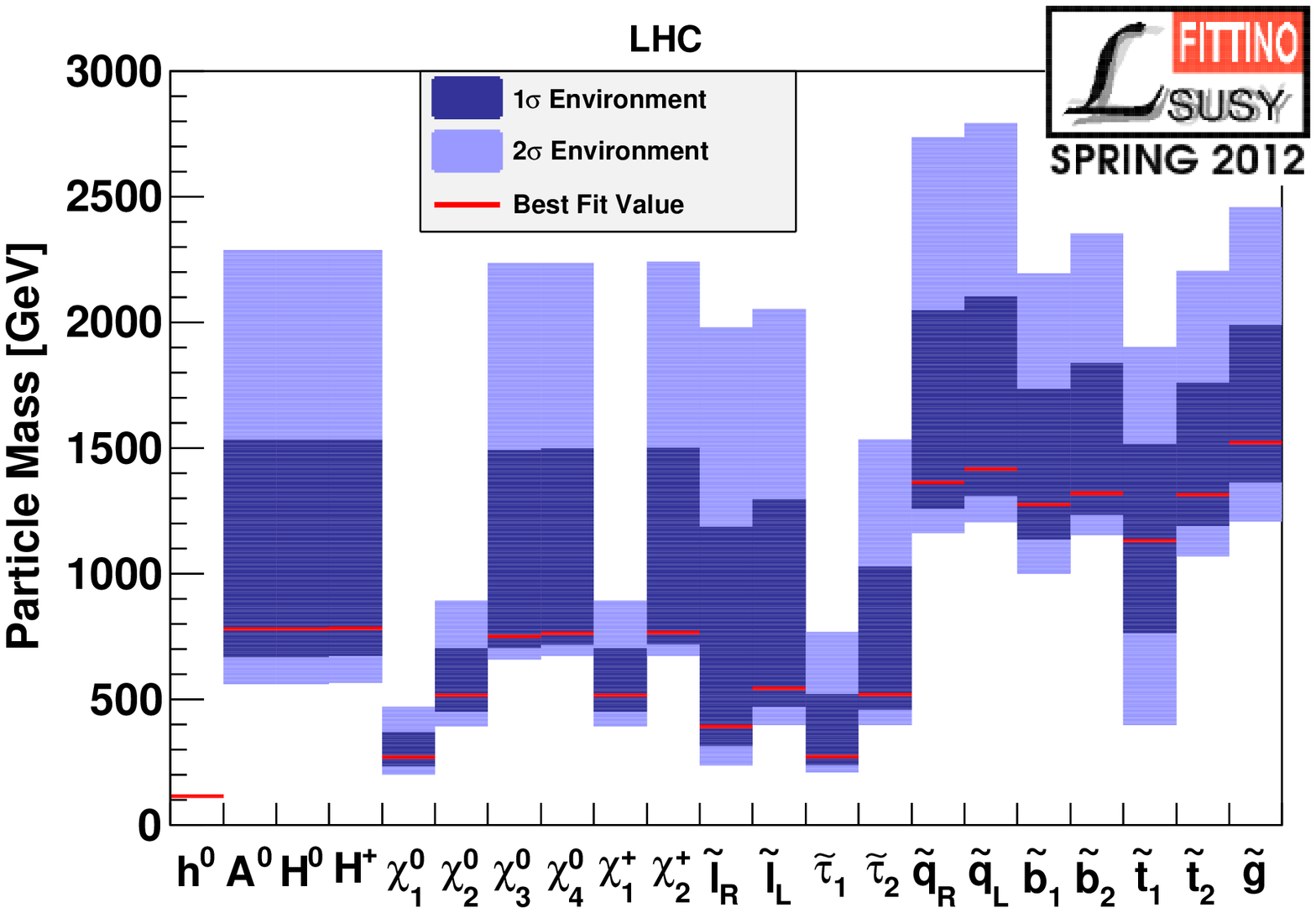}
\caption{Left: Parameter distributions for the LHC fit of the CMSSM with the 1-dimensional $1\sigma$ in red and the 2-dimensional $2\sigma$ in blue, and the best fit point marked by a star. 
Right: predicted distribution of sparticle and Higgs boson masses from the LHC fit of the CMSSM. The full uncertainty band gives the 1-dimensional $2\sigma$ uncertainty of each mass $\Delta\chi^2 < 4$.}
\label{fig:LHC}
\end{figure}

\begin{figure}
\includegraphics[width=.45\textwidth]{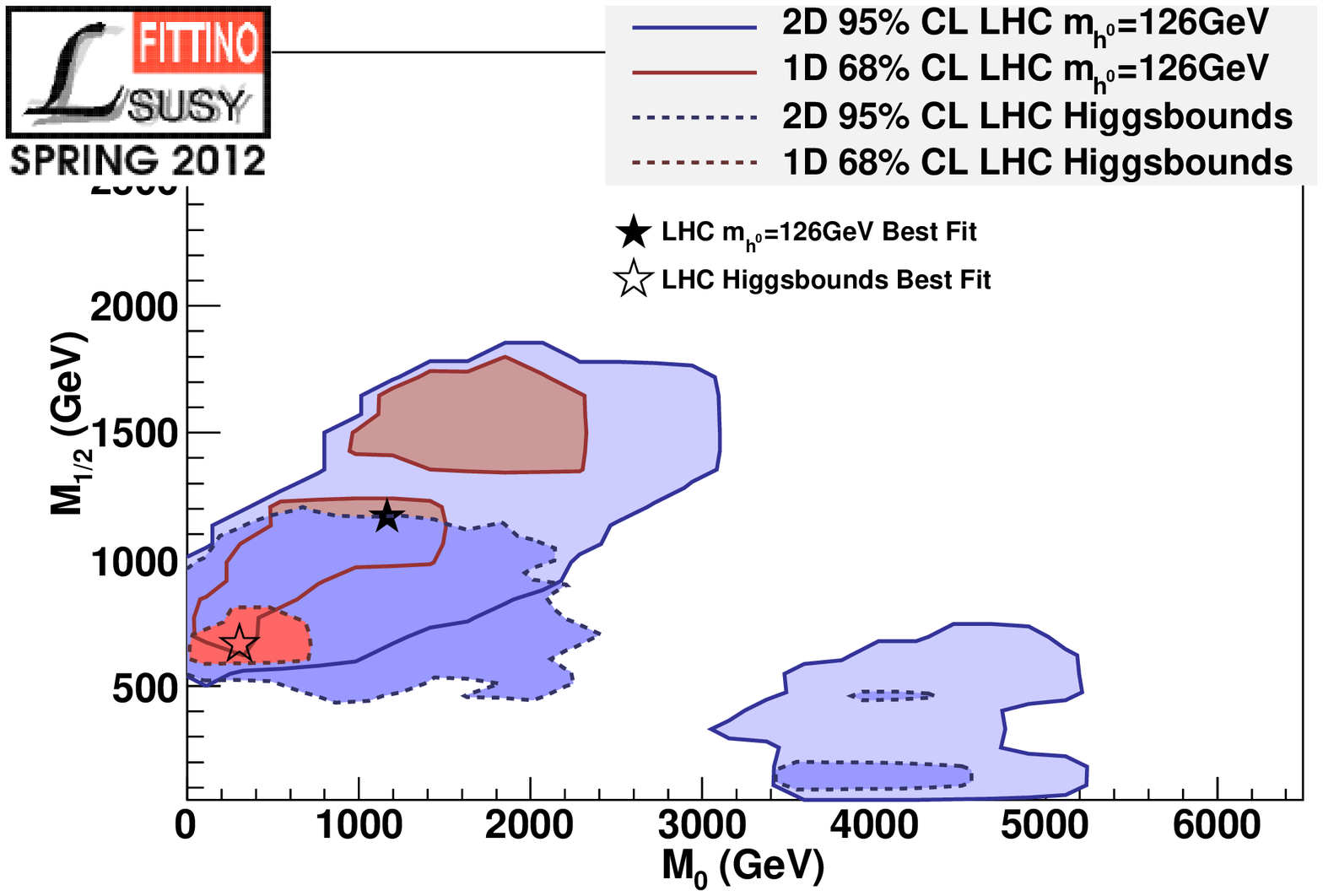}
\includegraphics[width=.45\textwidth]{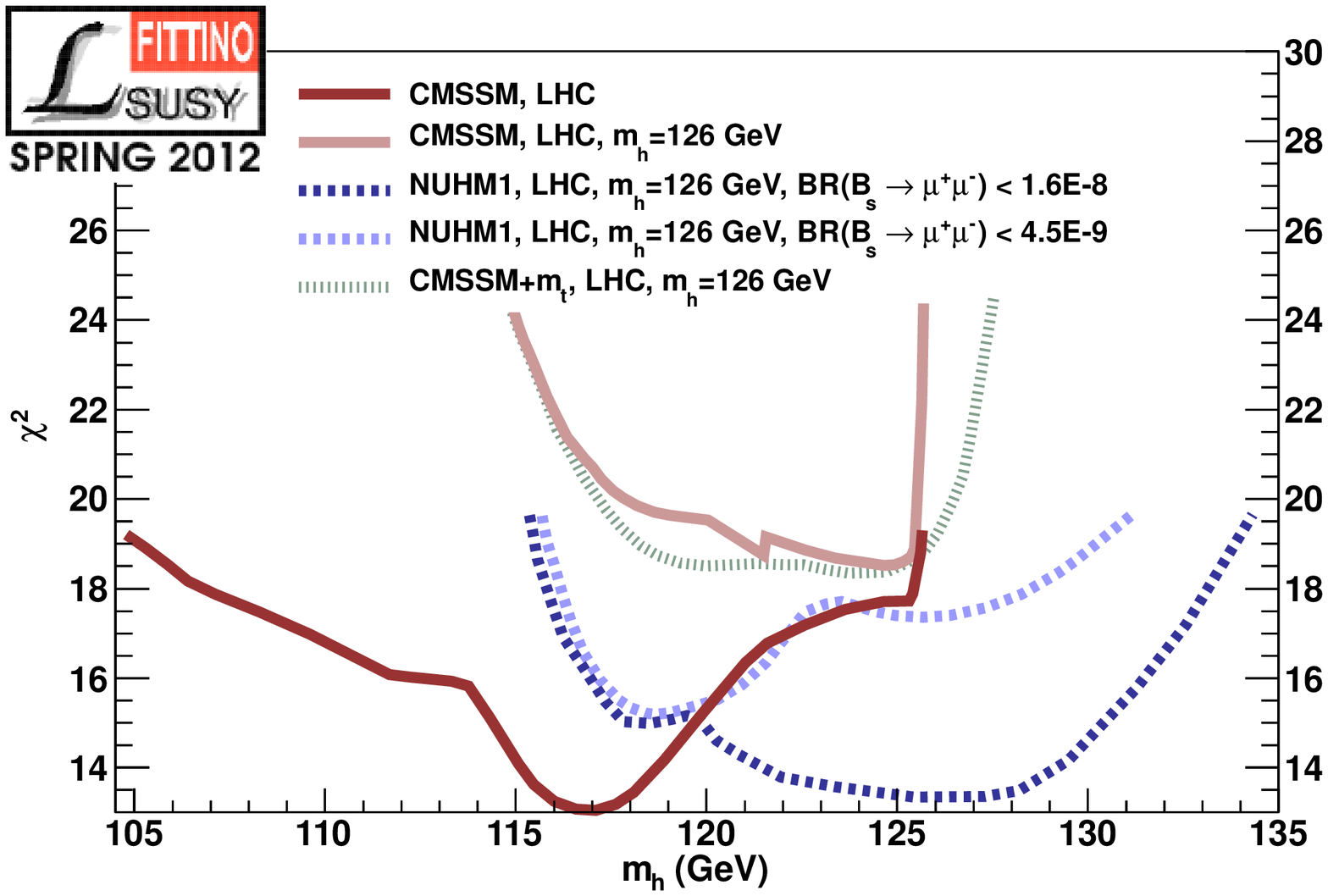}
\caption{Left: Parameter distributions for the LHC$+m_h$ fit of the CMSSM with the 1-dimensional $1\sigma$ in red and the 2-dimensional $2\sigma$ in blue, and the best fit point marked by a star. Right: The dependence of the minimal $\chi^2$ of the fit on $m_h$ for different input observable sets and for the CMSSM and NUHM1.}
\label{fig:H126}
\end{figure}

\begin{figure}
\includegraphics[width=.45\textwidth]{markovChain2D2sContours_P_M05_P_M1_.eps}
\includegraphics[width=.45\textwidth]{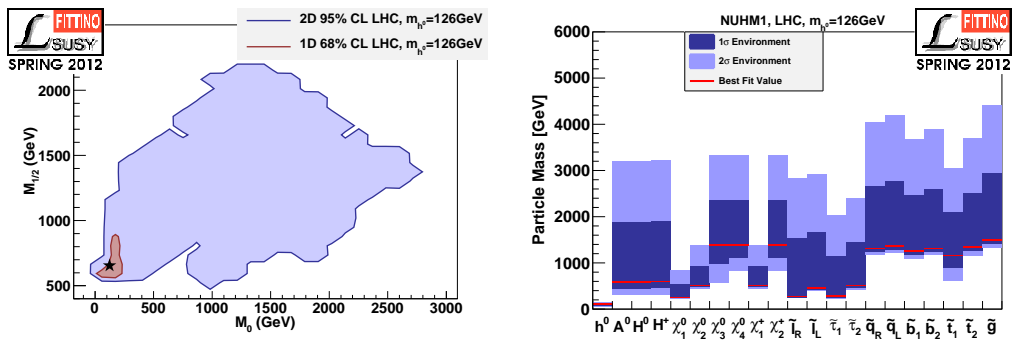}
\caption{Left: Parameter distributions for the LHC$+m_h$ fit of the NUHM1 with the 1-dimensional $1\sigma$ in red and the 2-dimensional $2\sigma$ in blue, and the best fit point marked by a star. Right: predicted distribution of sparticle and Higgs boson masses from the LH$C+m_h$ fit of the NUHM1.} 
\label{fig:NUHM}
\end{figure}

The results of our CMSSM and NUHM1 fits for various sets of input observables are given in the Table~\ref{tab:fitsummary}, for all fits we require the lightest neutralino to be the LSP, consistent radiative electroweak symmetry breaking and the absence of tachyons. In the plots of fitted parameters, the best-fit point is
marked by a star, all hidden dimensions having been profiled. Excluding the direct searches of super-symmetry at the LHC, the fitted parameter space points to a light sparticle spectrum, below 1~TeV (see Fig.~\ref{fig:LEO})). The focus point, {\it i.e.} $M_0\sim 2~$TeV and $M_{12}\sim 150$~GeV, is allowed in the $2\sigma$ region, due to the high Higgs mass $m_{h}$ prefered by current data. Including the direct search at the LHC decreases the goodness of the fit, which arises from the coupling of colored and non colored sectors in the constrained models. The low energy observables are indeed mainly driven by non colored sparticles, whilst the channels used for the direct search at the LHC rely mostly on the colored one. This results in a shift upwards of $\tan\beta$ and of the masses of squarks and gluinos (see Fig.~\ref{fig:LHC}), the shift in $\tan\beta$ being cause by the correlation with the masses through the muon anomaly: an increase in the mass being compensated by a larger coupling in order to match the experimental high value of $a_{\mu}$. Contrary to the CMSSM, the NUHM1 can accomodate a Higgs boson as heavy as 126~GeV, depending on the value of $BF(B_s\rightarrow\mu^+\mu^-)$ (see Fig.~\ref{fig:H126}). The fit of the NUHM1 model, including all experimental constraints, results in a large $2\sigma$ contour with a prefered region at lower mass and the focus point to be excluded (see Fig.~\ref{fig:NUHM}). Despite a lower fit tension than for the MSSM, a tension remains due to the strong correlation between $BF(B_s\rightarrow\mu^+\mu^-)$, $a_{\mu}$ and $m_{h}$ in the NUHM1.

\begin{table}[t]
  \renewcommand{\arraystretch}{1.5}
  \caption{Summary of the results for the CMSSM and NUHM1 fits with
    different sets of input observables: "LEO" refers to all low energy observables, "LHC" includes also the direct search for sparticles at the LHC, and "LHC+$m_h$" adds the constraint of a Higgs boson of 126~GeV.
  }\label{tab:fitsummary}
  \begin{center}
    \begin{tabular}{ccccccc}
      Fit & $M_0\, {\rm [GeV]}$ & $M_{1/2}\, {\rm [GeV]}$ & $M_H^2\,{\rm [10^6GeV^2]}$ & $\tan\beta$ & $A_0$ & $\chi^2$/$ndf$ \\ \hline\hline
      CMSSM: LEO & $84.4^{+144.6}_ {-28.1}$  & $375.4^{+174.5}_{ -87.5}$ & $\times$ & $14.9^{+16.5}_{-7.2}$ & 
      $186.3^{+831.4}_{-843.7} $ & 10.3/8 \\
      CMSSM: LHC & $304.4^{+373.7}_{-185.2}$ & $664.6^{+138.3}_{-70.9}$  &  $\times$ & $34.4^{+10.9}_{-21.3}$ &
      $884.8^{+1178.0}_{-974.9}$ & 13.1/9 \\
      CMSSM: LHC+$m_h$ & $1163.2^{+1185.3}_{-985.7}$  &
      $1167.4^{+594.0}_{-513.0}$  &  $\times$ & $39.3^{+16.7}_{-32.7}$  & $-2969.1^{+6297.8}_{-1234.9}$ & 18.4/9 \\
      NUHM1: LHC+$m_h$ & $124.3^{+95.2}_{-16.8}$  &
        $655.5^{+218.0}_{-65.0}$  & $-1.7^{+0.5}_{-2.7}$ & $29.4^{+3.3}_{-7.8}$  & $-511.2^{+574.7}_{-988.6}$ & 15.3/8 \\
    \end{tabular}
  \end{center}
\end{table}

When the constraint of a Higgs mass of $m_{h}=126$~GeV is included, the prefered values of masses and $\tan\beta$ increases (see Fig.~\ref{fig:H126}), while the quality of the fit decreases further, leading to a tension in the fit of constrained models. This tension can hardly be relieved by leaving the top mass free in the fit.

\begin{figure}
\includegraphics[width=.45\textwidth]{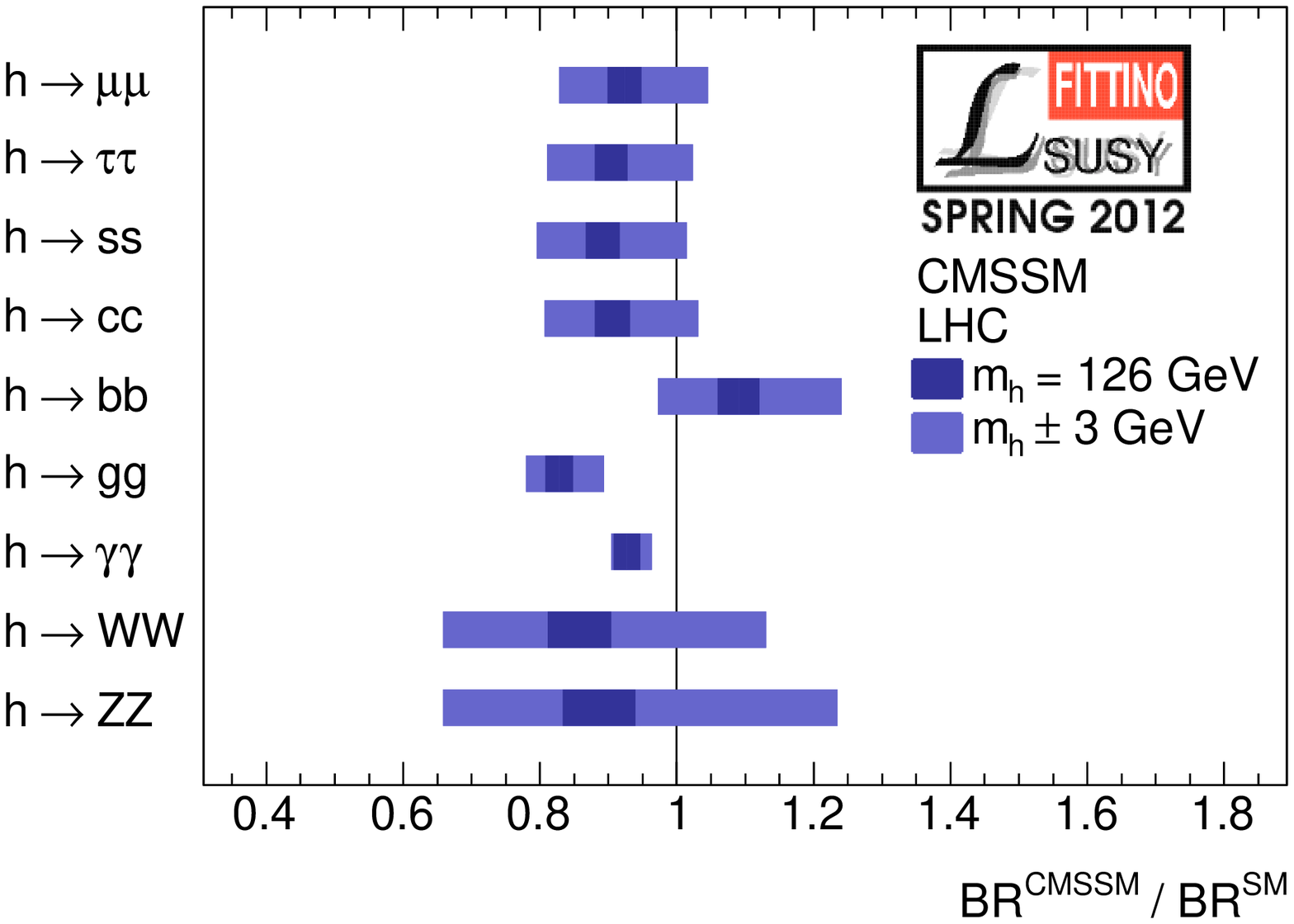}
\includegraphics[width=.45\textwidth]{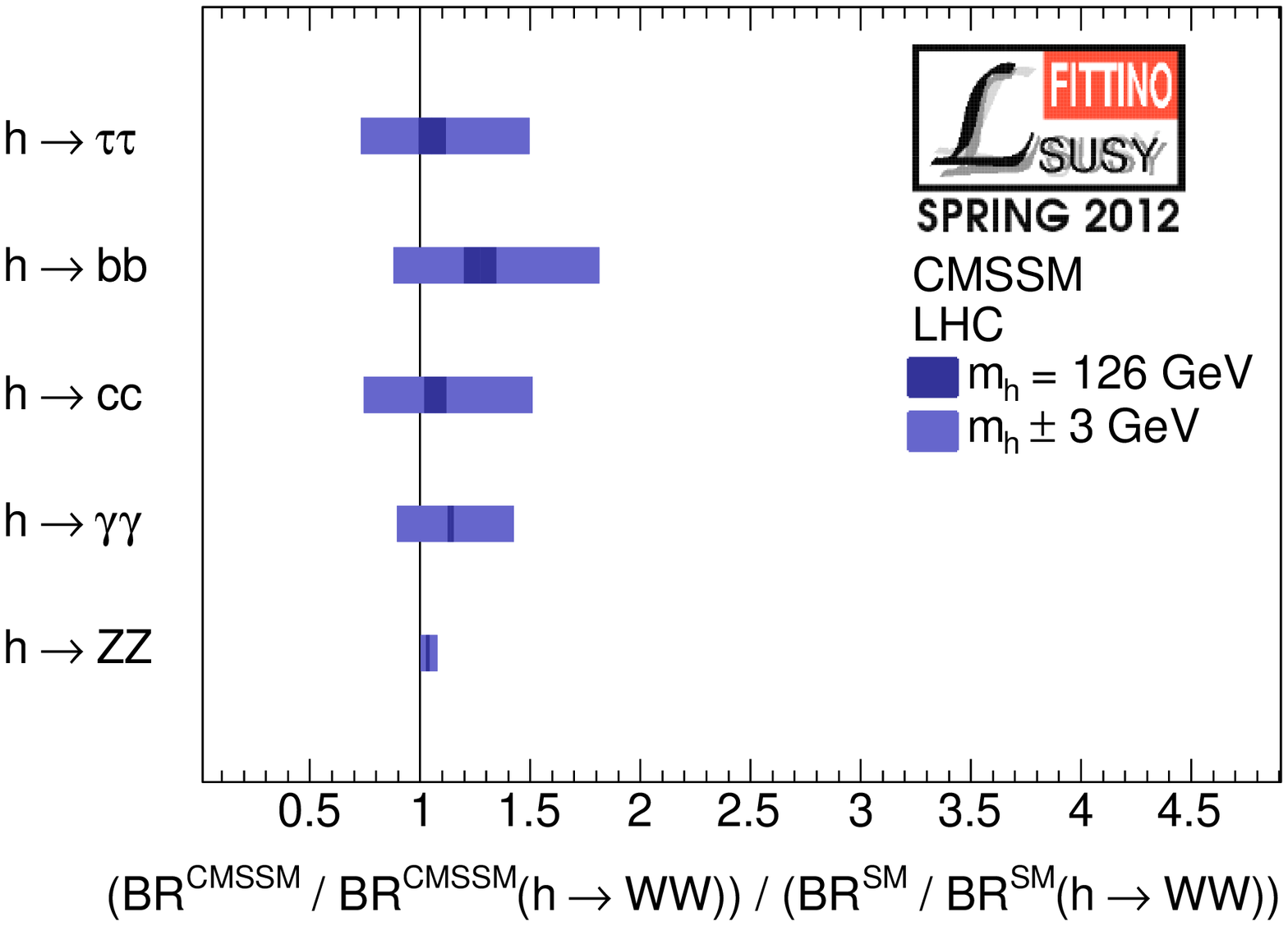}
\caption{Predicted $2\sigma$ ranges of Higgs branching fractions and ratios for the LHC fit of the CMSSM with $m_h=(126\pm 2\pm 3)$~GeV, including the theoretical uncertainty of the Higgs mass of 3~GeV.}
\label{fig:BR}
\end{figure}

Various ratios between the CMSSM and the SM of branching fractions of the main Higgs decay channels have been calculated with HDECAY 4.41~\cite{HDECAY} for the regions prefered by the fit in the parameter space (see Fig.~\ref{fig:BR}). One notices an enhancement of the $h\rightarrow b\bar{b}$ channel and a decrease of $h\rightarrow \tau^+\tau^-$ by regards to the SM.
Such a sensitivity makes potential measurements of the branching fractions attractive to discover a deviation from the SM and to determine the model parameters, even for SUSY mass scales beyond the LHC reach at $\sqrt s=7$~TeV or 8~TeV.

\section{Conclusion}

We presented a global frequentist fit of the CMSSM and NUHM1 parameter spaces, including up-to-date measurements in the flavor and electroweak sectors, the muon anomaly, astrophysical observations, the direct searches of supersymmetry at LHC and a Higgs mass of $m_{h}=126$~GeV. The current LHC exclusion leads to a low goodness-of-fit within the CMSSM, which worsens when requiring a Higgs mass above 125~GeV. The fit quality is increased in the non-minimal models NUHM1, despite a remaining tension due to the high correlations between observables in such model. The measurements of the Higgs branching fractions of the channels $h\rightarrow b\bar{b}$ and $h\rightarrow \tau^+\tau^-$ has the potential to indicate the deviation from the SM, even for SUSY mass scales beyond the LHC reach.

\end{document}